# Layer-dependent properties of SnS$_2$ and SnSe$_2$ novel two-dimensional materials

Joseph M. Gonzalez and Ivan I. Oleynik[*]

*Department of Physics, University of South Florida, Tampa, FL 33620*

The layer dependent structural, electronic and vibrational properties of SnS$_2$ and SnSe$_2$ are investigated using first-principles density functional theory (DFT). The in-plane lattice constants, interlayer distances and binding energies are found to be layer-independent. Bulk SnS$_2$ and SnSe$_2$ are both indirect band gap semiconductors with $E_g = 2.18$ eV and $1.07$ eV, respectively. Few-layer and monolayer 2D systems also possess an indirect band gap, which is increased to 2.41 eV and 1.69 eV for single layers of SnS$_2$ and SnSe$_2$. The effective mass theory of 2D excitons, which takes into account the combined effect of the anisotropy, non-local 2D screening and layer-dependent 3D screening, predicts strong excitonic effects. The binding energy of indirect excitons in monolayer samples, $E_x \sim 0.9$ eV, is substantially reduced to $E_x = 0.14$ eV in bulk SnS$_2$ and $E_x = 0.09$ eV in bulk SnSe$_2$. The layer-dependent Raman spectra display a strong decrease of intensities of the Raman active $A_{1g}$ mode upon decreasing the number of layers down to a monolayer, by a factor of 7 in the case of SnS$_2$ and a factor of 20 in the case of SnSe$_2$ which can be used to identify number of layers in a 2D sample.

## I. INTRODUCTION

The layered metal dichalcogenides, consisting of chemically inert layers bonded together by weak van der Waals (vdW) interactions represent an emerging class of two-dimensional (2D) materials beyond graphene[1–3]. The unique electronic properties of single and few layer samples of these materials are being explored to develop novel applications in electronics[4], photonics[5], chemical sensing[6], catalysis[7] and energy storage[8] to mention a few. Much of recent research activities in this field has been focused on transition metal chalcogenides such as MoS$_2$[9–11], WS$_2$[9,12,13], MoSe$_2$[14], MoTe$_2$[15] and TaS$_2$[16], while the interest in other 2D $s$-$p$ metal chalcogenides has emerged only recently[17–21]. The discovery of the direct band gap in single layer of MoS$_2$[5], which possesses the indirect band gap in the bulk[5] sparked the excitement of 2D research community, resulting in the observation of several unusual phenomena including very interesting exciton physics in these 2D materials[22–25]. In this regard, one of the important questions is whether other layered indirect band gap semiconductors display a similar transformation when the dimensionality is reduced from 3D in the bulk to 2D single or few layer samples.

One of the representative examples of layered $s$-$p$ metal chalcogenides are group IV-VI semiconductors tin disulfide (SnS$_2$) and tin diselenide (SnSe$_2$). These compounds possess a layered structure and exhibit a rich polytypism which results from the various stacking sequences of identical S-Sn-S (Se-Sn-Se) layers. SnS$_2$ and SnSe$_2$ have been extensively studied in their bulk form in the past[26–33]. Recently, these single and few layer materials have been employed in several applications including phase change memory[7], water splitting[20], field-effect transistors[34], gas sensing[35], and high-speed photodetection[21,36]. The weakly-bonded layered structure of these compounds allows one to use traditional exfoliation techniques to isolate single and few layers of SnS$_2$[7,18,37]. In addition, these 2D materials can be grown using van-der-Waals epitaxy[16,37,38], vapor transport[39], molecular beam epitaxy[14] and chemical vapor deposition (CVD)[21,39]. Depending on the growth conditions, SnS$_2$ can appear in several polytypes including 2H, 4H and 18R, while SnSe$_2$ exists in 2H and 18R polytypes[40]. Although the atomic structure of the monolayer of SnS$_2$ or SnSe$_2$ is unambiguously defined, it is the different stacking sequence of such layers that produces an unique crystal structure of such polytypes. In this work, the 2H polytype is used as the bulk material from which the few-layer 2D samples of SnS$_2$ and SnSe$_2$ are produced.

Recently, Huang *et. al.*[18] experimentally investigated few-layer samples of 4H polytype of SnS$_2$ and reported the first experimental measurements of the band structure of a monolayer of SnS$_2$. It was found that the indirect to direct band gap transition does not occur as the single layer still possesses an indirect band gap. In addition, they also observed a monotonic decrease in the intensity of the $A_{1g}$ Raman mode with decreasing number of layers and suggested that Raman spectroscopy can be effectively used to determine the number of layers in the sample. The group fabricated several field effect transistor devices using samples of varying thickness and demonstrated the increase in carrier mobility in few layer samples compared to that measured in bulk crystals[18].

Although a single layer of SnS$_2$ has been considered by several theoretical[17,41] and experimental[18] groups, the few-layer material properties of both SnS$_2$ and SnSe$_2$ have not been systematically studied so far. Therefore, the goal of this work is to investigate the evolution of the structural, electronic and vibrational properties of SnS$_2$ and SnSe$_2$ 2D materials as a function of the number of layers (i.e. going from bulk down to a single layer) using first-principles density functional theory (DFT). Several computational challenges such as a proper description of weak van der Waals (vdW) interactions and the deficiency of traditional DFT to predict the band gaps of semiconductors are addressed in this work by using the DFT-D2 vdW empirical correction proposed by Grimmie[42] for calculation of the atomic structure and the hybrid functional of Heyd, Scuseria, and Ernzerhof (HSE06)[43] – for investigation of the electronic structure. The robust performance of the HSE06 hybrid functional in describing electronic properties is validated by comparing the HSE06 electronic properties for the bulk crystals with those obtained by using the quasi-particle GW method. The layer-dependent properties such as band gaps, effective masses, dielectric constants, exciton binding energies, and Raman spectra are then calculated and compared with the available experimental data.

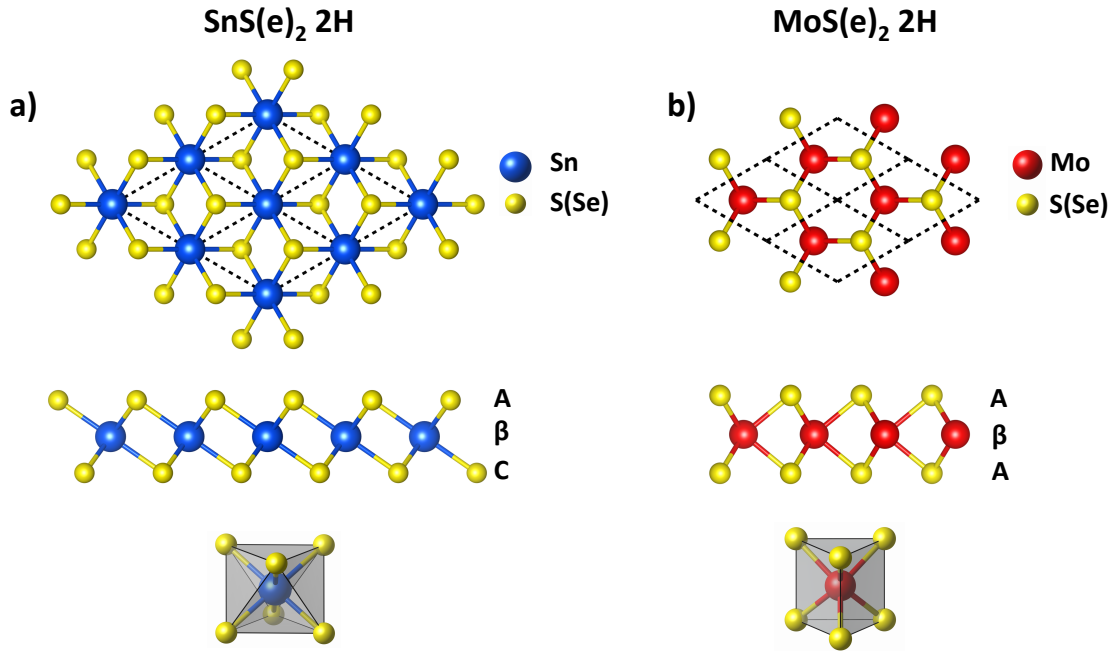

FIG. 1. (Color online) Atomic structure for a monolayer of (a) SnS$_2$ (SnSe$_2$) and (b) MoS$_2$ (MoSe$_2$), demonstrating a very different local atomic environment of the metallic atoms in both compounds.

## II. COMPUTATIONAL DETAILS

First-principles calculations are carried out using Vienna *ab initio* simulation package (VASP)[44]. Due to the importance of vdW interactions in weakly-bonded layered compounds, and given the inability of standard DFT to properly describe these long range dispersive interactions, the performance of several standard and state-of-the-art functionals are evaluated. Specifically, the standard local density approximation (LDA), the Perdew, Burke, and Ernzerhof (PBE) generalized gradient approximation (GGA)[45], the vdW-DF2 functional of Langreth et al.[46], and the empirical correction DFT-D2 by Grimmie[42], are used to calculate the ground state lattice parameters of the bulk SnS$_2$ and SnSe$_2$ crystals and compared with experiment. The atomic structure of the crystals is determined by optimizing the atomic positions using the conjugate-gradient method with a force convergence criterion of $10^{-2}$ eV/Å. A cutoff energy of 400 eV, and a $\Gamma$-centered Monkhorst-Pack[47] k-point grid with a density of 0.03 Å$^{-1}$ ($11 \times 11 \times 6$ grid) are employed in the atomic and electronic structure calculations. This choice of kpoint grid and energy cutoff are chosen such that the maximum force on each atom is less than 0.03 eV/Å and the stresses are converged to within 0.1 GPa. For all layered structures, the Monkhorst-Pack grid is $11 \times 11 \times 1$.

The electronic properties are calculated using the HSE06 hybrid functional with the same plane wave cutoff and k-point grid used in the structural relaxation. The band structure plots are obtained by interpolating the energies using the maximally-localized Wannier functions provided by the WANNIER90[48] code. The contribution of the spin-orbit interactions has been evaluated and found to be vanishingly small. Therefore the spin orbit coupling is not included in layer-dependent studies of electronic structure presented in this work. A vacuum slab of 20 Å along the $c$ axis is used in calculations to avoid spurious interaction between periodic images. This set of parameters were chosen such that the band energies are converged to within 18 meV.

The vibrational properties of SnS$_2$ and SnSe$_2$ 2D materials in this work include off-resonant Raman frequencies and corresponding activities (intensities). This requires the calculation of phonons at the $\Gamma$-point and corresponding derivatives of the macroscopic dielectric tensor with respect to the normal mode coordinates. These quantities are computed within the frozen phonon approximation as described in the paper by Porezag and Pederson[49]. The calculations are performed using the PBE GGA functional with D2 empirical vdW corrections with a k-point grid density of $0.02\,\text{Å}^{-1}$ ($16 \times 16 \times 9$ Monkhorst-Pack grid) and a plane-wave energy cutoff of 800 eV. For all layered structures, the k-point grid is $16 \times 16 \times 1$. The increase of both the energy cutoff and kpoint grid are necessary to achieve the high accuracy calculation of the forces, resulting in the determination of vibrational frequencies converged to within $2\,\text{cm}^{-1}$.

Table I. Lattice parameters of bulk $SnS_2$ and $SnSe_2$ crystals calculated by different density functionals including those accounting for weak van der Waals interactions between layers. Experimental values are listed for comparison. Shown in parentheses is the error between calculated and experimental values.

| System | | LDA | PBE | DF2 | PBE+D2 | Expt.[31] |
|--------|------|------|------|------|--------|-----------|
| $SnS_2$ | a (Å) | 3.63(-0.27%) | 3.70(1.65%) | 3.83(5.22%) | 3.68(1.09%) | 3.64 |
| | c (Å) | 5.69(-3.23%) | 6.88(17.01%) | 6.04(2.72%) | 5.89(0.17%) | 5.88 |
| | c/a | 1.57(-3.09%) | 1.86(14.81%) | 1.58(-2.47%) | 1.60(-1.23%) | 1.62 |
| $SnSe_2$ | a (Å) | 3.80(-0.26%) | 3.87(1.57%) | 4.04(6.04%) | 3.83(0.52%) | 3.81 |
| | c (Å) | 5.90(-3.91%) | 6.96(13.36%) | 6.34(3.26%) | 6.17(0.49%) | 6.14 |
| | c/a | 1.55(-3.73%) | 1.80(11.80%) | 1.57(-2.48%) | 1.61(0%) | 1.61 |

## III. STRUCTURAL PROPERTIES

There exists a wide-spread confusion in labeling the crystal structures of different classes of metal chalcogenide compounds. The $sp$-metal chalcogenides of $CdI_2$-type are usually labeled using Ramsdell notation[50], listing the number of chalcogen atomic sheets in the unit cell followed by a letter (H - for hexagonal, R - for rhombohedral, or C - for cubic lattice types), e.g. 2H is the label for a simple $CdI_2$-type crystal structure consisting of a single $A\beta C$ layer periodically repeated in $c$-direction (capital roman letters depict the stacking close-packed (111) positions of the chalcogens whereas greek letter in between - stacking position of the metal atom). The 2H polytype has space group of $P\bar{3}m1$ and octahedral coordination of the metal atom surrounded by chalcogen atoms, see Fig. 1(a). The most common polytype of $SnS_2$ and $SnSe_2$ is 2H. Unfortunately, the same 2H label is used for the ground state crystal structure of $MoS_2$ where the Mo metal atom possesses trigonal prismatic coordination, see Fig. 1(b). In addition, the 2H labeled unit cell contains two S-Mo-S layers (hence 2 in front of the letter), not a single layer as in the case of $SnS_2$. In this work, we adopt the Ramsdell notation, which is widely used in the case of $CdI_2$-type crystals and build the few layer systems from the 2H bulk crystals.

The lattice parameters of bulk $SnS_2$ and $SnSe_2$, calculated by different DFT functionals, including vdW DF2 and D2, are listed in Table I and compared with experimental data. The LDA underestimates the out-of-plane lattice constant $c$ by 3.2% for $SnS_2$ and by 3.9% for $SnSe_2$, whereas in-plane lattice constant $a$ is predicted fairly accurately within 0.2% for both materials. The PBE functional overestimates the in-plane lattice constant $a$ by

1.6-1.7 %, whereas the out-of-plane lattice constant $c$ is poorly predicted, the errors being 17% ($SnS_2$) and 13.4% ($SnSe_2$). The vdW-DF2 method, while improving upon GGA-PBE, still appreciably overestimates the interlayer lattice constant $c$ by $\sim 3\%$ for both $SnS_2$ and $SnSe_2$, while the in-plane lattice parameter is overestimated by nearly 6%, which is unsatisfactory. In contrast, Grimmie PBE+D2 method provides an excellent description of the crystal structure of the studied materials, with an average error less than 1% for both $a$ and $c$ lattice constants. The ability of PBE+D2 to reproduce in-plane lattice parameters close to experimental values is important as the artificial in-plane strain influences the value of the band gap[9]. Therefore, the monolayer and few-layered structures of $SnS_2$ and $SnSe_2$ are obtained by cleaving (0001) layers of bulk crystals relaxed by PBE+D2 method. After cleaving, the layered structures are relaxed. It is found that the in-plane lattice parameters of few layer systems are largely preserved, a maximum deviation from the bulk values being $0.01\,\text{Å}$.

The interlayer binding energy per unit area $E_b$ of $n$-layer system ($n \neq 1$) is calculated as $E_b = (nE_1 - E_n)/(nA)$ where $E_1$ and $E_n$ are the energies of monolayer and $n$-layer systems, and $A$ is the surface area of the 2D unit cell $\sim 12\,\text{Å}^2$. Using this definition, the interlayer binding energy is found to be $13.1\,\text{meV/Å}^2$ for bi-layer compared to $13.5\,\text{meV/Å}^2$ for the bulk in case of $SnS_2$; and $18.4\,\text{meV/Å}^2$ for bi-layer compared to $18.9\,\text{meV/Å}^2$ for the bulk in case of $SnSe_2$. These results demonstrate that the structural properties of both $SnS_2$ and $SnSe_2$ 2D materials are nearly layer-independent. The interlayer binding energies are within the range of values reported for other layered chalcogenides[51].

## IV. ELECTRONIC PROPERTIES

It is well known that both the LDA and GGA functionals underestimate the values of band gaps for most insulators and semiconductors, while the GW approximation and hybrid functionals predict band gaps in much closer agreement with experiment. To demonstrate the necessity to go beyond standard DFT, the band struc-ture of 2H polytypes of bulk $SnS_2$ and $SnSe_2$ crystals are calculated using the PBE GGA functional, the HSE06 hybrid functional and the GW approximation, see Fig. 2. Upon close inspection of Fig. 2, the topologies of the PBE, GW and HSE06 bands are very similar. In particular, for both $SnS_2$ (Fig. 2(a)) and $SnSe_2$ (Fig. 2(b)) the three methods predict nearly identical valence band topologies, with the PBE functional underestimating the fundamental band gap relative to the HSE06 and GW methods.

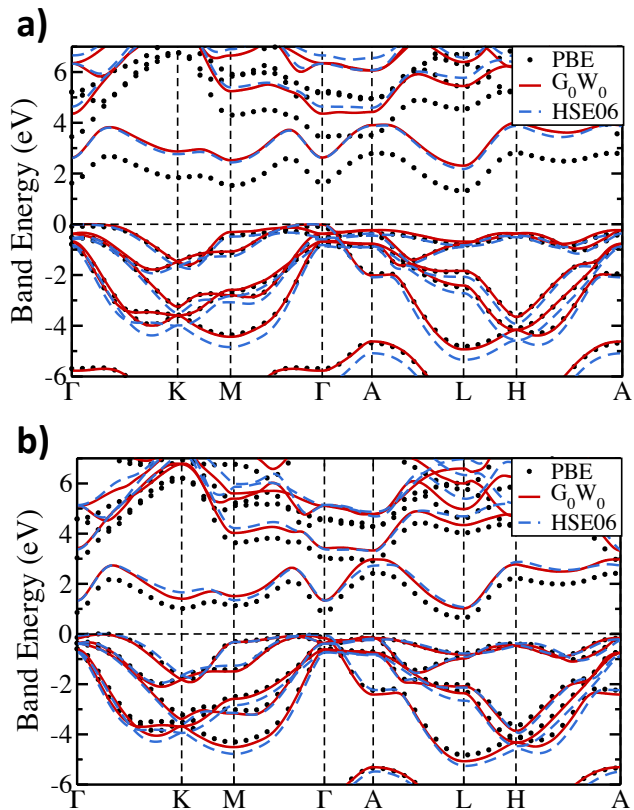

FIG. 2. Band structure of bulk (a) $SnS_2$ and (b) $SnSe_2$ calculated by three methods: PBE GGA functional (black dots), the $G_0W_0$ approximation (red line) and the hybrid functional HSE06 (blue dashed line).

Although the GW and HSE results are in good agreement with each other, the high computational cost of the GW method suggests the use of HSE06 hybrid functional, which provides a good balance between accuracy and computational expense. Therefore, all results presented below are obtained using the HSE06 hybrid functional. As mentioned in the computational details section, the spin-orbit interactions are found to be negligible, i.e. for both compounds the splitting of the top valence band is $\approx 13\,\text{meV}$.

It can clearly be seen from Fig. 2, that both bulk $SnS_2$ and $SnSe_2$ crystals possess an indirect band gap, with the valence band maxima (VBM) located along the line $\Gamma - M$, see Fig. 3. In addition, the conduction band minimum is located at the L point for both compounds, see Fig. 2. The indirect band gap for $SnS_2$ is calculated to be $E_g^{in} = 2.18\,\text{eV}$ and the minimum energy of direct $L \rightarrow L$ transition is $E_g^{dir} = 2.61\,\text{eV}$ , see Fig. 2(a).

Table II. Calculated indirect ($E_g^{in}$) and direct ($E_g^{dir}$) fundamental band gaps of bulk and few-layer $SnS_2$ and $SnSe_2$ materials. Bulk values calculated at the $L$ high symmetry point and layered values calculated at the $M$ high symmetry point. Experimental band gaps for the bulk $SnS_2$ and $SnSe_2$ are listed in parentheses.

| System | | $E_g^{in}$ (eV) | $E_g^{dir}$ (eV) |
|---|---|---|---|
| $SnS_2$ | Bulk | 2.18(2.18[28],2.28[52]) | 2.61(2.88[28],2.56[52]) |
| | 4-Layer | 2.22 | 2.50 |
| | 3-Layer | 2.29 | 2.54 |
| | 2-Layer | 2.34 | 2.57 |
| | 1-Layer | 2.41 | 2.68 |
| $SnSe_2$ | Bulk | 1.07( 0.98[28],1.06[53]) | 1.84(1.62[28],1.28[53]) |
| | 4-Layer | 1.26 | 1.58 |
| | 3-Layer | 1.37 | 1.68 |
| | 2-Layer | 1.51 | 1.83 |
| | 1-Layer | 1.69 | 2.04 |

The band structure of bulk $SnSe_2$ shown in Fig. 2(b), is very similar to that of $SnS_2$, with the only major difference being a lower magnitude of both indirect and direct band gaps. The calculated indirect band gap for $SnSe_2$ was determined to be $E_g^{in} = 1.07\,\text{eV}$ and the direct gap, due to $L \rightarrow L$ transition is $E_g^{dir} = 1.84\,\text{eV}$ , see Fig. 2(b). This is in good agreement with available experimental data, see Table II. The theoretical band gaps of both $SnS_2$ and $SnSe_2$ bulk crystals structures can be directly compared with those obtained from optical measurements since their reduction due to exciton binding energies $E_x$ is expected to be small, see the next section for discussion of excitonic effects.

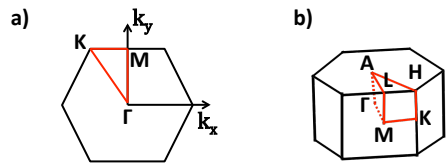

FIG. 3. Schematics of (a) two-dimensional, and (b) three-dimensional Brillouin zones.

The layer dependence of the band structure of $SnS_2$ is shown in Fig. 4(a). In contrast to $MoS_2$, $SnS_2$ remains an indirect band-gap semiconductor even in monolayer form. Similar to the bulk band structure, the VBM is at a point along the $\Gamma - M$ line for single and few layer systems, the only difference being a change in the degree of concavity around this point. In all few-layered structures, the conduction band minimum (CBM) is located at the $M$ high symmetry point, which is equivalent to the $L$ high symmetry point in the bulk Brillouin zone, see Fig. 3. The indirect band gaps, $E_g^{in}$, are calculated to be 2.18 eV for the bulk and 2.41 eV for the monolayer with intermediate values for the few-layer systems, see Fig. 5. For few layer systems, the direct band gaps $E_g^{dir}$ occur at the $M$ point of the 2D Brillouin zone, whereas for the bulk it occurs at the $L$ point of the 3D Brillouin zone, from which $E_g^{dir}$ is calculated to be 2.68 eV for the monolayer and 2.61 eV for the bulk. Although $E_g^{dir}$ is reduced upon increasing the number of layers, it does not converge to the bulk value of 2.61 eV within the first four layers. This is because of appreciable dispersion of electronic bands along $z$ direction of the Brillouin zone resulting in a measurable difference between the band energies at $L$ and $M$ points of the 3D Brillouin zone, see Fig. 4(a). The increase of the band gaps in the layered structures upon reduction of number of layers can be attributed to the effective reduction of the screening of electrostatic interactions in few-layer systems surrounded by vacuum[54–56] as well as quantum confinement of electrons within a quasi-two dimensional material of finite thickness[10].

Table III. Layer-dependence of the electron $m_e^*(k_x)$, $m_e^*(k_y)$ and hole $m_h^*(k_x)$, $m_h^*(k_y)$ effective masses in the $k_x$ and $k_y$ direction of the Brillouin zone for SnS$_2$ and SnSe$_2$. Also listed are the 2D reduced effective masses, $\mu_i = (m_i^e m_i^h)/(m_i^e + m_i^h)$, the exciton 2D and 3D reduced effective masses, $\mu^{2D} = 2\left(\mu_x^{-1} + \mu_y^{-1/3}\mu_z^{-2/3}\right)^{-1}$ and $\mu^{3D} = 3\left(\mu_x^{-1} + \mu_y^{-1} + \mu_z^{-1}\right)^{-1}$ which are in units of the electron rest mass. The calculated screening length, $r_0$ and the exciton binding energies, $E_x$ are also given.

| System | | $m_e^*(k_x)$ | $m_e^*(k_y)$ | $m_h^*(k_x)$ | $m_h^*(k_y)$ | $\mu_x$ | $\mu_y$ | $\mu_x/\mu_y$ | $\mu^{2D}$ | $\mu^{3D}$ | $r_0$(Å) | $E_x$(eV) |
|---|---|---|---|---|---|---|---|---|---|---|---|---|
| | 1-Layer | 0.342(M) | 0.815(M) | 0.342 | 2.266 | 0.171 | 0.599 | 0.285 | 0.206 | - | 18.158 | 0.912 |
| | 2-Layer | 0.364(M) | 1.023(M) | 0.373 | 2.474 | 0.184 | 0.724 | 0.255 | 0.226 | - | 7.041 | 0.304 |
| SnS$_2$ | 3-Layer | 0.401(M) | 1.618(M) | 0.419 | 2.263 | 0.205 | 0.943 | 0.217 | 0.256 | - | 6.913 | 0.238 |
| | 4-Layer | 0.391(M) | 1.718(M) | 0.398 | 2.252 | 0.197 | 0.975 | 0.202 | 0.249 | - | 7.083 | 0.201 |
| | Bulk | 0.375(L) | 1.104(L) | 0.424 | 2.542 | 0.199 | 0.770 | 0.259 | - | 0.345 | - | 0.137 (0.112[35]) |
| | | | | | | | | | | | | |
| | 1-Layer | 0.348(M) | 0.811(M) | 0.354 | 2.188 | 0.175 | 0.592 | 0.297 | 0.256 | - | 21.299 | 0.855 |
| | 2-Layer | 0.377(M) | 0.826(M) | 0.371 | 2.284 | 0.187 | 0.607 | 0.308 | 0.269 | - | 8.338 | 0.229 |
| SnSe$_2$ | 3-Layer | 0.381(M) | 1.314(M) | 0.386 | 2.027 | 0.192 | 0.797 | 0.241 | 0.282 | - | 8.184 | 0.168 |
| | 4-Layer | 0.388(M) | 1.158(M) | 0.369 | 2.347 | 0.189 | 0.775 | 0.244 | 0.278 | - | 8.407 | 0.142 |
| | Bulk | 0.425(L) | 1.107(L) | 0.502 | 2.228 | 0.230 | 0.739 | 0.311 | - | 0.363 | | 0.093 |

The layer dependence of the electronic band structure for SnSe$_2$, shown in Fig. 4(b), is similar to that of SnS$_2$: the VBM is at a point located along the line $\Gamma - M$, and the CBM occurs at the $M$ point. Additionally, the indirect band gap, $E_g^{in}$, is increased upon reduction of the layer thickness down to a monolayer, i.e. $E_g^{in}$ changes from 1.07 eV for the bulk to 1.69 eV for the monolayer with a similar trend observed for the direct band gap, see Fig. 5(b). However, similar to the case of SnS$_2$, $E_g^{dir}$ does not converge to the bulk value within the first four layers, see Table II. As explained above, this is because $E_g^{dir}$ for few layer and bulk samples are measured at different points of the 2D and 3D Brillouin zones. For both SnS$_2$ and SnSe$_2$, the transformation of the band structure from the bulk down to monolayer causes a slight flattening of the top valence bands resulting in an increase in the density of states at the valence band edge, while the lower-energy bands remain largely unchanged. There is also a slight change in the topology of the lowest conduction bands upon reducing the number of layers. The values of the effective masses obtained by fitting along the $k_x$ and $k_y$ directions of the Brillouin zone are presented in Table III. For both SnS$_2$ and SnSe$_2$, the monolayer samples possess the lightest electron effective mass, although not by a large margin. The effective electron and hole masses are found to be slightly higher for the SnSe$_2$ system, which is in accordance with a similar trend, $m_S^* < m_{Se}^*$, observed in the transition metal family of chalcogenides[9]. However, there is no clear trend in the dependence of the effective masses on the number of layers."

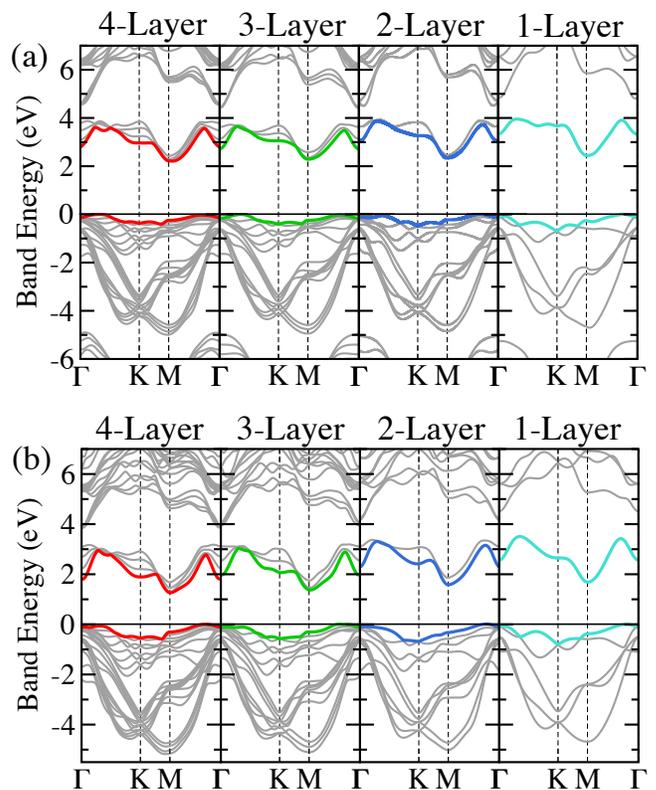

FIG. 4. Layer-dependent electronic band structure of (a) SnS$_2$ and (b) SnSe$_2$. (Color online) The thick colored lines are highlighted to showcase the band edge.

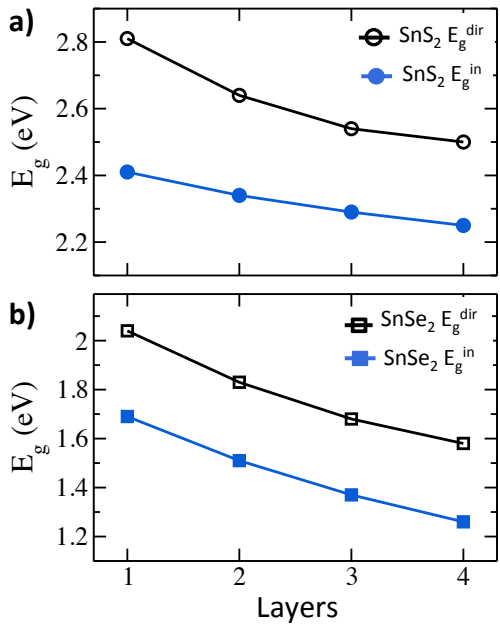

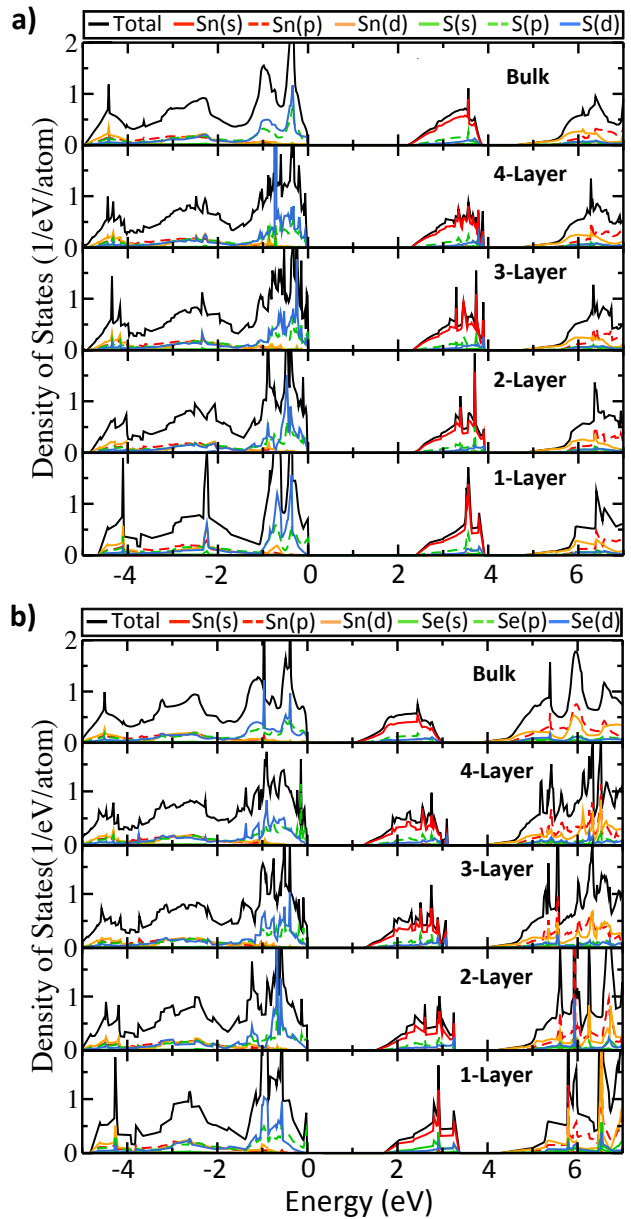

FIG. 5. Layer dependent indirect $\left(E_g^{in}\right)$ and direct $\left(E_g^{dir}\right)$ energy band gaps for SnS$_2$ and SnSe$_2$.

The calculated layer-dependent total and local density of states (LDOS) for SnS$_2$ and SnSe$_2$ are used to analyze the orbital contributions to the electronic structure, see Fig. 6. For both bulk and few layered structures, the valence band edge is dominated by chalcogen (S or Se) $p$ and $d$ orbitals. However, the conduction band edge is dominated by Sn $s$ orbitals with a small contribution from the chalcogen S (or Se) $p$ orbitals. One notable feature at the valence band edge is an increase of the DOS at this energy upon decreasing the number of layers in the sample. This increase in DOS is consistent with the flattening of the bands near the top of the valence band. In addition, both monolayer and bilayer DOS exhibit several characteristic peaks due to van Hove singularities in two dimensions. In the range of energies greater than 4 eV the LDOS is dominated by Sn $p$ and $d$ orbitals with very little contributions from any of the chalcogen orbitals.

## V. EXCITONS IN LAYERED STRUCTURES

The optical excitations in semiconductors are substantially influenced by electron-hole interactions[57]. In bulk crystals, the exciton binding energy is typically on the order of tens of meVs, which is much smaller than the fundamental band gap. This is because of a substantial screening of the electron-hole Coulomb interactions by the dielectric medium of the crystal as well as small electron and hole effective masses at the bottom and the top of the conduction and valence bands, respectively. In contrast, the excitonic effects are significantly amplified due to a combined effect of quantum and dielectric confinements in 2D materials. The quantum confinement is due to the reduced dimensionality of the quantum-mechanical Hamiltonian, which results in a factor of four increase of exciton binding energies over the standard 3D Hydrogenic model. The dielectric confinement is due to the reduced screening of electron-hole interactions of few-layer systems surrounded by vacuum, which results in an additional increase of the exciton binding energy by up to a factor of ten. The dielectric screening in 2D is non-local, i.e. the e-h interaction is strongly screened at short distances resulting in a weak logarithmic dependence of the potential, whereas it is a mostly unscreened $1/r$ dependent Coulomb interaction at large distances $r$. The cross-over distance is the screening length $r_0$, which is determined via 2D polarizability $\chi_{2D}$: $r_0 = 2\pi\chi_{2D}/\kappa_{eff}$, where $\kappa_{eff}$ is the effective dielectric constant of the environment surrounding the 2D system.

FIG. 6. Total and local density of states for bulk and few-layered structures for (a) SnS$_2$ and (b) SnSe$_2$.

In our work, we adopt the effective mass theory of 2D excitons developed by Velizhanin et al[58], which provides a physically transparent description of the effects of the non-local screening, and the reduced dimensionality, which is not too dissimilar to 2D effective mass theories developed recently by other researchers[12,56,59–64]. Within this model, the exciton binding energy $E_x$ is a function of the screening length $r_0$, the exciton's reduced effective mass $\mu$, which is in units of the electron rest mass, and the effective dielectric constant of the environment surrounding the 2D layer $\kappa_{eff}$:

$$E_x = 4\frac{\mu}{\kappa_{eff}^2}\mathcal{E}\left(\frac{r_0}{a}\right)E_{Ry}, \qquad (1)$$

where the effective Bohr radius $a = a_0\kappa_{eff}/\mu$, $a_0 = 0.53$ Å, $E_{Ry} = 13.65$ eV. $\mathcal{E}(r_0/a)$ is the dimensionless universal function of the normalized screening length, $r_0/a$, which describes the effect of the non-local screening on the binding energy of an exciton and has been calculated using path integral Monte Carlo method, see Ref.[58] for tabulated values of $\mathcal{E}$. $\mathcal{E}(r_0/a_0)$ monotonically decreases from 1 at $r_0 = 0$ (no screening) to smaller values at large $r_0/a$, reflecting the fact that the screening effectively reduces the electron-hole electrostatic interactions, thus making $E_x$ smaller.

The effective dielectric constant $\kappa_{eff}$ takes into account an additional screening of electron-hole interactions in few layer samples, see Eq. (1). Specifically, the individual layers in the interior of the sample experience the bulk dielectric environment, whereas the top and the bottom boundary layers experience both vacuum from the outside and the bulk dielectric environment from the interior. The average effective dielectric constant is then $\kappa_{eff} = ((N-1)\kappa_B + 1)/N$, where $N$ is the number of layers and $\kappa_B = \sqrt{\epsilon_z\epsilon_{xy}}$, where $\epsilon_{xy}$ is the in-plane component, and $\epsilon_z$ the transverse component of the bulk dielectric tensor of the bulk crystal, which was calculated within the random phase approximation (RPA)[65]. The values of $\kappa_B$ are calculated to be 5.85 for SnS$_2$ and 7.32 for SnSe$_2$. For a single layer, $\kappa_{eff} = 1$, since it is surrounded by vacuum on both sides.

The screening length, $r_0$, for a specific few-layer sample is determined by calculating the 2D polarizability $\chi_{2D}$, which is obtained by fitting the dependence of the in-plane dielectric constant, $\epsilon_{xy}$, on the thickness, $L$, of the vacuum layer separating periodic images of the few-layer system[63], $\epsilon_{xy}(L) = 1 + 4\pi\chi_{2D}/L$ . The 2D screening is reduced with increasing $\kappa_{eff}$ since $r_0 \propto 1/\kappa_{eff}$, thus providing the transition to the functional form of a pure 3D Coulomb interaction $1/r$ at large $\kappa_{eff}$. Both $\chi_{2D}$ and $r_0$ are plotted as a function of number of layers in Fig. 7(a) for SnS$_2$ and in Fig. 7(b) for SnSe$_2$. Although $\chi_{2D}$ monotonically increases upon increasing the number of layers, the effective screening length $r_0$ displays an opposite trend because of the increase in $\kappa_{eff}$.

The few-layer samples as well as the bulk samples of SnS$_2$ and SnSe$_2$ exhibit a substantial in-plane anisotropy of the electron and hole effective masses, see Table III. Therefore, the approach of Velizhanin et al[58] is modified by introducing the reduced effective exciton mass defined as $\mu^{2D} = 2\left(\mu_x^{-1} + \mu_y^{-1/3}\mu_x^{-2/3}\right)^{-1}$, where $\mu_x$ and $\mu_y$ are the corresponding $x$ and $y$ components of electron and hole reduced masses: $\mu_i = (m_i^e m_i^h)/(m_i^e + m_i^h)$, $i = \{x,y\}$. This expression for $\mu^{2D}$ was obtained by Prada et al[61] by finding a variational solution for the anisotropic 2D exciton. The bulk exciton binding energies are calculated by neglecting anisotropy in all three directions as the anisotropy factor $\gamma = \epsilon_{xy}\mu_{xy}/(\epsilon_z\mu_z)$ is close to 1 in both cases of SnS$_2$ and SnSe$_2$. Therefore, the bulk exciton binding energy is calculated as $E_x = \mu^{3D}E_{Ryd}/\kappa_B^2$ with $\mu^{3D} = 3\left(\mu_x^{-1} + \mu_y^{-1} + \mu_z^{-1}\right)^{-1}$, where $\mu_x$ and $\mu_y$ are reported in Table III and $\mu_z = 0.42$ for SnS$_2$ and $\mu_z = 0.39$ for SnSe$_2$.

The calculated exciton binding energies for few layer and bulk samples are listed in the Table III and plotted in Fig. 7. The monolayer exciton binding energies $E_x$ are substantial, 0.91 eV and 0.86 eV for SnS$_2$ and SnSe$_2$, respectively. The values of $E_x$ for bi-layer are reduced by more than a factor of three due to the combined effects of reduced 2D dielectric confinement (due to the decrease of the screening length $r_0$ and corresponding increase of $\mathcal{E}(r_0/a)$) and the onset of 3D screening (the increase of $\kappa_{eff}$), see Eq. (1). Upon further increase of the number of layers, the exciton binding energies are approaching those in the bulk SnS$_2$, 0.14 eV, and SnSe$_2$, 0.09 eV. To our knowledge, there exists no information on the exciton binding energy in the bulk SnSe$_2$, however in the case of SnS$_2$, the calculated value of $E_x$ for the bulk exciton is in reasonable agreement with experiment, $E_x = 0.11$ eV[35].

As compared to the transition metal dichalcogenides MoS$_2$ and WS$_2$[12,58,66], the monolayer exciton binding energies $E_x$ in $s$-$p$ metal chalcogenides SnS$_2$ and SnSe$_2$ are greater by a factor of two despite $E_x$ for the bulk structures being comparable for all four compounds: $E_x \sim 0.1$ eV. In addition, the exciton effective masses for both bulk and monolayer samples are very similar as well: $\mu \sim 0.1 - 0.3$. The enhancement of $E_x$ can be explained by a markedly different 2D polarizability in these two classes of materials. For example, single layers of MoS$_2$ and WS$_2$ have a 2D polarizability of $\sim 6.5$ Å, while single layers of SnS$_2$ and SnSe$_2$ have a polarizability of $\sim 3.5$ Å. The corresponding 2D screening length, $r_0 = 2\pi\chi_{2D}$, quantifies the reduction of 2D dielectric screening in both SnS$_2$ and SnSe$_2$ compared to MoS$_2$ and WS$_2$, resulting in smaller values of $\mathcal{E}$ in Eq. (1). The markedly different monolayer 2D polarizabilities can be explained by a substantial difference in the atomic po-larizabilities of the constituent metallic atoms[67]: 6.28 Å³ for Sn compared to h 12.8 Å³ for Mo and 11.2 Å³ for W. The larger polarizabilities of the transition metal atoms can be attributed to the presence of $d$-orbitals which are more loosely bound than the $p$-orbitals in Sn. This correlation between the enhancement of the 2D excitonic binding energies and the atomic polarizabilities is also seen in the case of phosporene, which also possesses a large 2D exciton binding energy of 0.90 eV[68] and small atomic polarizability 3.63 Å³.

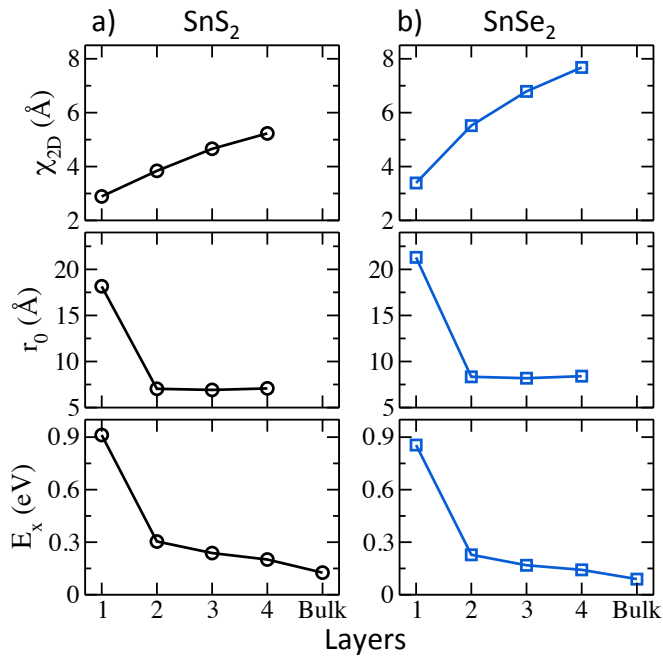

FIG. 7. Layer dependent (a) Two dimensional polarizability, $\chi_{2D}$, (b) screening length, $r_0$, and (c) exciton binding energy, $E_x$.

## VI. VIBRATIONAL PROPERTIES

Raman spectroscopy is a widely used to characterize 2D materials[69–71] including the identification of the number of layers in 2D samples[18,71]. To aid in the interpretation of experiments, the evolution of the Raman spectrum for both $SnS_2$ and $SnSe_2$ is investigated as the number of layers is reduced from infinite (bulk) to single layer, see Fig. 8.

The 2H polytype for $SnS_2$ and $SnSe_2$ contains three atoms in the unit cell, therefore, there are nine normal modes, three of which are Raman-active. The lowest frequency mode $E_g$, is doubly degenerate and is characterized by an in-plane stretching behavior shown in Fig. 9(a). The other non-degenerate higher frequency Raman mode $A_{1g}$ is characterized by an out of plane stretching of the chalcogen atoms, see Fig. 9(b). The $A_{1g}$ displays a much larger intensity than the lower frequency $E_g$ mode. For example, the calculated intensity of the $A_{1g}$ peak of the bulk $SnS_2$ is $\sim 7$ times weaker than that of $E_g$ peak, see Fig. 8.

Table IV. Layer-dependent Raman frequencies for the bulk and few layer samples of $SnS_2$ and $SnSe_2$. The available experimental frequencies are given in parentheses.

| System | | $E_g$ (cm⁻¹) | $A_{1g}$ (cm⁻¹) |
|---|---|---|---|
| | Bulk | 205.1(205.5[32]) | 310.8(315.5[32]) |
| | 4-Layer | 206.1 | 309.5 |
| $SnS_2$ | 3-Layer | 206.2 | 308.9 |
| | 2-Layer | 207.3 | 307.0 |
| | 1-Layer | 206.1(200[18]) | 304.6(315[18]) |
| | | | |
| | Bulk | 119.4(115.5[19]) | 191.0(188.3[19]) |
| | 4-Layer | 118.7 | 188.5 |
| $SnSe_2$ | 3-Layer | 117.5 | 187.9 |
| | 2-Layer | 114.5 | 186.5 |
| | 1-Layer | 108.3 | 184.1 |

In the case of bulk $SnS_2$, the calculated Raman frequencies of the $E_g$ mode, 205.1 cm⁻¹ and $A_{1g}$ mode, 310.8 cm⁻¹, are in good agreement with experimentally measured frequencies 205.5 cm⁻¹ and 313.5 cm⁻¹ reported by Smith *et. al.*[32]. For the monolayer of $SnS_2$, the calculated Raman frequencies 191.6 cm⁻¹ ($E_g$ mode) and 304.6 cm⁻¹ ($A_{1g}$ mode)

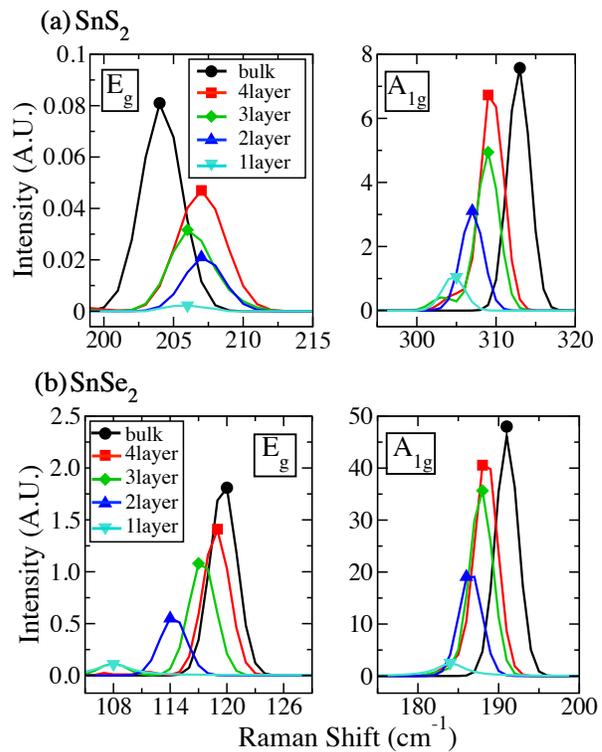

FIG. 8. Raman spectra of 2H bulk and few layer samples of of (a) $SnS_2$ and (b) $SnSe_2$. The atomic vibrations corresponding to the $E_g$ and $A_{1g}$ Raman modes are depicted in Fig. 9

are also in good agreement with the experimental frequencies of the peaks at 200.0 cm⁻¹ and 315.0 cm⁻¹ re-ported by Huang *et. al.*[18], the typical error being characteristic of frequency underestimation by DFT. For the bulk SnSe$_2$, the calculated frequencies of the $E_g$ and $A_{1g}$ modes, 119.4 cm$^{-1}$ and 191.0 cm$^{-1}$, agree well with those from experiment, 115.5 cm$^{-1}$ and 188.3 cm$^{-1}$, reported by Taube *et. al.*[19]. To our knowledge, there are no reports on single or few layer Raman spectra for SnSe$_2$. Although the frequencies of the $A_{1g}$ mode display a monotonic reduction as the number of layers changes from infinite (the bulk) down to a monolayer for both SnS$_2$ and SnSe$_2$, the situation is not so certain in the case of $E_g$ mode. Although its frequency also decreases in the case of SnSe$_2$, for SnS$_2$ this trend is not observed, with the frequencies being somewhat layer-independent.

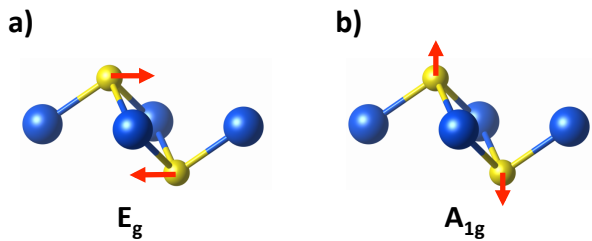

FIG. 9. Schematic of atomic displacements of the chalcogen atoms in Raman-active vibrational modes (a) $E_g$ and (b) $A_{1g}$ for the 2H polytype of SnS$_2$ and SnSe$_2$.

For both materials, the intensities of the $E_g$ and $A_{1g}$ modes experience a strong reduction with decreasing layers, see Fig. 8. For example the intensity of the $A_{1g}$ mode in a monolayer of SnSe$_2$ is $\sim$ 19x less than the intensity of the corresponding peak in the bulk crystal.

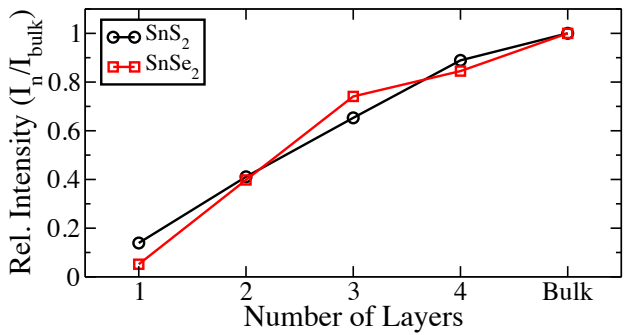

FIG. 10. Relative intensity of the $A_{1g}$ Raman mode in SnS$_2$ and SnSe$_2$ as a function of the number of layers.

Even though the intensity of $E_g$ mode is much smaller than that of $A_{1g}$ mode, both of them display an intensity reduction upon decrease of number of layers from infinite (the bulk) to a monolayer. In fact, the mono-tonic decrease of the intensity of the $A_{1g}$ mode, shown for both SnS$_2$ and SnSe$_2$ in Fig. 10, can be used to identify the number of layers in the sample, which was first suggested in the experimental work of Huang *et. al.*[18]. This strong layer dependence of the intensity for the $A_{1g}$ mode is reasonable given the associated atomic displacements, see Fig. 9(b), i.e. with fewer layers, the effect of the concerted motion out of the plane is less substantial, while with more layers, this motion is amplified.

## VII. CONCLUSIONS

In summary, we have investigated layer-dependent properties of novel 2D materials, SnS$_2$ and SnSe$_2$. It was found that the structural properties of both SnS$_2$ and SnSe$_2$, including in-plane lattice parameters, interlayer distances and binding energies are nearly layer-independent. The electronic structure calculations using the HSE06 hybrid functional demonstrate that the nature of the indirect band gap does not change when reducing the number of layers from infinite in the bulk down to a monolayer for both SnS$_2$ and SnSe$_2$.

The monolayers of SnS$_2$ and SnSe$_2$ display strong excitonic effects, which are studied by applying a novel effective mass theory of 2D excitons developed by Velizhanin *et al*[58], modified to include the effect of anisotropy in effective masses. The monolayer binding energies of indirect excitons $E_x \sim 0.9$ eV are substantially reduced to $E_x = 0.14$ eV, and $E_x = 0.09$ eV for bulk SnS$_2$ and SnSe$_2$ respectively.

The layer-dependent Raman spectra for both SnS$_2$ and SnSe$_2$ display only a weak increase of the frequencies of the $A_{1g}$ and $E_g$ Raman active modes upon increase of number of layers, whereas their intensities display dramatic increase by a factor of 7 and 20, respectively. This strong layer dependence of Raman intensities can be used as a practical means of counting the number of layers in 2D samples.

The predicted strong layer dependence of electronic, excitonic and vibrational properties of 2D SnS$_2$ and SnSe$_2$ materials suggest new experiments which can provide new insights into fundamental properties of 2D materials.

## ACKNOWLEDGMENTS

Simulations were performed using the NSF XSEDE facilities (grant No. TG-MCA08X040), the USF Research Computing Cluster supported by NSF (grant No. CHE-1531590), and the computational facilities of the Materials Simulation Laboratory at USF.

---

* oleynik@usf.edu